\documentclass[aps,prl,twocolumn,letterpaper,superscriptaddress]{revtex4} %preprint
\usepackage{dsfont}
\usepackage{amsmath}
\usepackage{graphicx}
\usepackage{multirow}
\usepackage[latin1]{inputenc}
\usepackage{color}
\usepackage{colortbl}
\usepackage{xcolor}  % for writting parts in color, e.g. in red

\usepackage[FIGTOPCAP, hang, raggedright, nooneline]{subfigure}
\usepackage{extarrows}
\usepackage{ulem}

\newcommand{\eq}[1]{(\ref{eq:#1})}
\newcommand{\Eq}[1]{Eq.~(\ref{eq:#1})}

\makeindex

\begin{document}

\title{Experimental Observation of a Generalized Gibbs Ensemble} 

\author{T. Langen} 
\email{tlangen@ati.ac.at}
\affiliation{Vienna Center for Quantum Science and Technology, Atominstitut, TU Wien, Stadionallee 2, 1020 Vienna, Austria}

\author{S. Erne}
\affiliation{Vienna Center for Quantum Science and Technology, Atominstitut, TU Wien, Stadionallee 2, 1020 Vienna, Austria}
\affiliation{Institut f\"ur Theoretische Physik,
             Ruprecht-Karls-Universit\"at Heidelberg,
             Philosophenweg~16,
             69120~Heidelberg, Germany}
\affiliation{ExtreMe Matter Institute EMMI,
             GSI, 
             Planckstra\ss e~1, 
             64291~Darmstadt, Germany} 
	
\author{R. Geiger}
\affiliation{Vienna Center for Quantum Science and Technology, Atominstitut, TU Wien, Stadionallee 2, 1020 Vienna, Austria}
								
\author{B. Rauer}
\affiliation{Vienna Center for Quantum Science and Technology, Atominstitut, TU Wien, Stadionallee 2, 1020 Vienna, Austria}

\author{T. Schweigler}
\affiliation{Vienna Center for Quantum Science and Technology, Atominstitut, TU Wien, Stadionallee 2, 1020 Vienna, Austria}
		
\author{M. Kuhnert} 
\affiliation{Vienna Center for Quantum Science and Technology, Atominstitut, TU Wien, Stadionallee 2, 1020 Vienna, Austria}
	
\author{W.~Rohringer}
\affiliation{Vienna Center for Quantum Science and Technology, Atominstitut, TU Wien, Stadionallee 2, 1020 Vienna, Austria}
												
\author{I. E. Mazets}
\affiliation{Vienna Center for Quantum Science and Technology, Atominstitut, TU Wien, Stadionallee 2, 1020 Vienna, Austria}
\affiliation{Wolfgang Pauli Institute, 1090 Vienna, Austria}
\affiliation{Ioffe Physico-Technical Institute, 194021, St. Petersburg, Russia}

\author{T. Gasenzer}
\affiliation{Institut f\"ur Theoretische Physik,
             Ruprecht-Karls-Universit\"at Heidelberg,
             Philosophenweg~16,
             69120~Heidelberg, Germany}
\affiliation{ExtreMe Matter Institute EMMI,
             GSI, 
             Planckstra\ss e~1, 
             64291~Darmstadt, Germany} 
						
\author{J. Schmiedmayer} 
\email{schmiedmayer@atomchip.org}
\affiliation{Vienna Center for Quantum Science and Technology, Atominstitut, TU Wien, Stadionallee 2, 1020 Vienna, Austria}

\begin{abstract}
The connection between the non-equilibrium dynamics of isolated quantum many-body systems and statistical mechanics is a fundamental open question. It is generally believed that the unitary quantum evolution of a sufficiently complex system leads to an apparent maximum-entropy state that can be described by thermodynamical ensembles. However, conventional ensembles fail to describe the large class of systems that exhibit non-trivial conserved quantities. Instead, generalized ensembles have been predicted to maximize entropy in these systems. In our experiments we explicitly show that a degenerate one-dimensional Bose gas relaxes to a state that can be described by such a generalized ensemble. This is verified through a detailed study of correlation functions up to 10th order. The applicability of the generalized ensemble description for isolated quantum many-body systems points to a natural emergence of classical statistical properties from the microscopic unitary quantum evolution.
\end{abstract}

\date{\today}

\maketitle
%------------------------------------

Information theory provides powerful concepts for statistical mechanics and quantum many-body physics. In particular, the principle of entropy maximization~\cite{Shannon:1949,Jaynes57,Jaynes57b} leads to the well-known thermodynamical ensembles, which are fundamentally constrained by conserved quantities like energy or particle number~\cite{HuangBook}. However, generic systems can contain many more conserved quantities, raising the question whether there exists a more general statistical description for the steady states of quantum many-body systems~\cite{Polkovnikov11}. 

Specifically, the presence of non-trivial conserved quantities puts constraints on the available phase space of a system, which strongly affects the dynamics~\cite{Gring12,Ronzheimer13,Caux13,Calabrese2011} and inhibits thermalisation~\cite{Kinoshita06,Rigol07,Rigol2009a}. Instead of relaxing to steady states described by the usual thermodynamical ensembles, a generalized Gibbs ensemble (GGE) was proposed to describe the corresponding steady states via the many-body density matrix~\cite{Jaynes57b,Rigol07,Cazalilla06,Caux12} 
\begin{equation}
\hat \rho = \frac{1}{Z}\exp\Big(-\sum_m\lambda_m\,\mathcal{\hat I}_m\Big).
\label{eq:gge}
\end{equation}
Here, $\mathcal{\hat I}_m$ denotes a set of conserved quantities and $Z=\mathrm{Tr}\exp(-\sum_m\lambda_m\mathcal{\hat I}_m)$ is the partition function. The Lagrange multipliers $\lambda_m$ associated with the conserved quantities are obtained by maximization of the entropy under the condition that the expectation values of the conserved quantities are fixed to their initial values. It is important to note that the emergence of such a maximum-entropy state is not in contradiction to a unitary evolution according to quantum mechanics. It rather reflects that the true quantum state is indistinguishable from the maximum-entropy ensemble with respect to a set of measurable observables~\cite{Polkovnikov11}.

%===========================================================================
\begin{figure}[tb]
	\centering
		\includegraphics[width=0.35\textwidth]{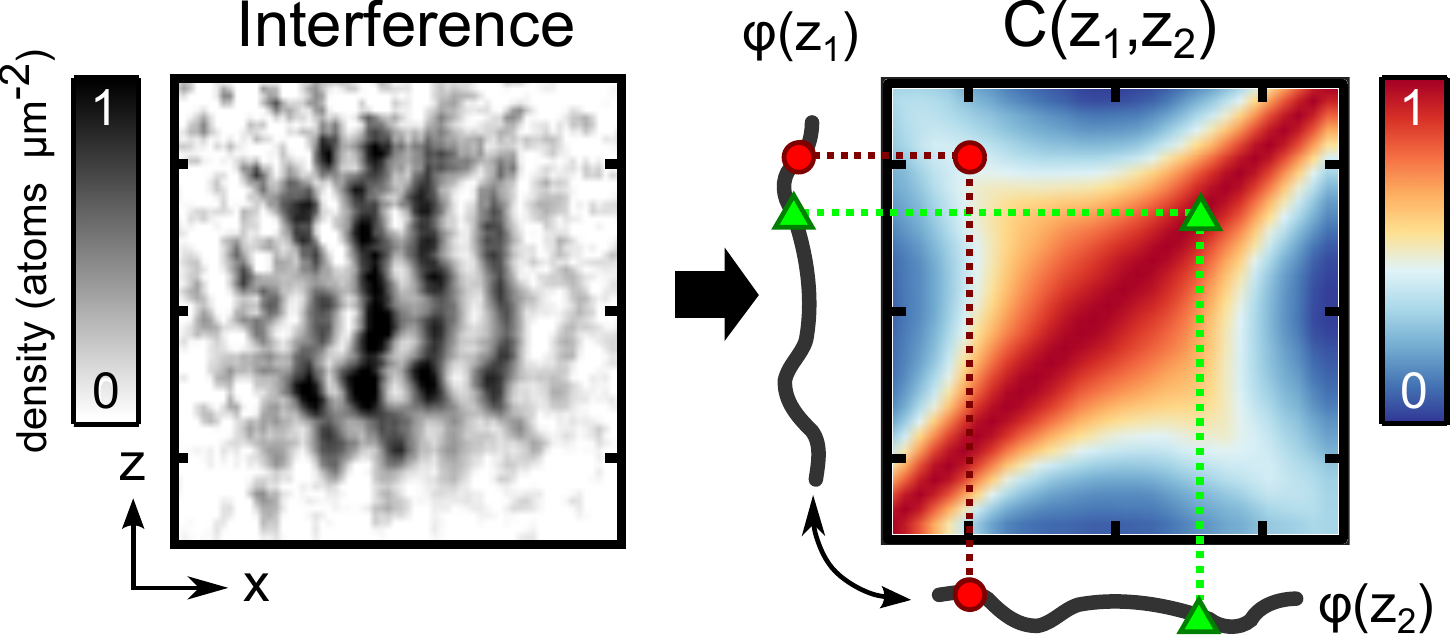}
	\caption{In the experiment, a non-equilibrium system is prepared by splitting a 1D Bose gas into two halves. After an evolution time $t$, matter-wave interference is used to extract the local relative phase profile $\varphi(z)$ between the two halves. This is accomplished by measuring the local position of the fluctuating interference fringes. Subsequently, the phase profile is used to calculate the two-point correlation function $C(z_1,z_2)\sim\langle \exp[i\varphi(z_1)-i\varphi(z_2)]\rangle$ as a function of all possible coordinates $z_1$ and $z_2$ along the length of the measured phase profile. For example, as every point is perfectly correlated with itself, all coordinates where $z_1=z_2$ (green triangles) lead to $C(z_1,z_2)=1$. These coordinates are located on the diagonal of the correlation function. As another example, coordinates with $z_1=-z_2$ (red points) are located symmetrically around the center of the system and found on the anti-diagonal of the correlation function.}
	\label{fig:Fig1}
\end{figure}
%===========================================================================

The GGE is a direct generalization of the usual thermodynamical ensembles and formally capable of describing a wide range of dynamically emerging steady states~\cite{Rigol08}. For example, in the case where only the energy is conserved, the GGE reduces to the standard canonical or Gibbs ensemble, with temperature being the only Lagrange multiplier~\cite{HuangBook}. Moreover, it famously provides a description for the steady states of integrable systems, which exhibit as many independent conserved quantities as they have degrees of freedom~\cite{Rigol07,Caux12,Calabrese2011}. Even if quantities are only approximately conserved, the GGE description was formally shown to be valid for significant time scales~\cite{Kollar11}. The GGE has also been suggested as a description for many-body localized states~\cite{Vosk2013}. However, while numerical evidence for the emergence of a GGE has been provided for many systems, a direct experimental observation has so far been lacking. 

Here, we experimentally study the relaxation of a trapped one-dimensional (1D) Bose gas. Our system is a close realization of the Lieb-Liniger model describing a homogeneous gas of 1D bosons with contact interactions, which is one of the prototypical examples of an integrable system~\cite{Lieb63,Korepin93}. In the thermodynamic limit, its exact Bethe Ansatz solutions imply an infinite number of conserved quantities, which make it impossible for the gas to forget an initial non-equilibrium state, forcing it to relax to a GGE. Recent results have shown that also the trapped 1D Bose gases that are realized in our and other experiments behave approximately integrable over very long time scales, enabling the detailed investigation of integrable dynamics~\cite{Kinoshita06,Gring12,Kuhnert13,Langen13b,Ronzheimer13}. To demonstrate the emergence of a GGE, we prepare such a 1D Bose gas in different initial non-equilibrium states and observe how they each relax to steady states that maximize entropy according to the initial values of the conserved quantities.

The experiments start with a phase-fluctuating 1D Bose gas~\cite{Petrov00} of ${}^{87}$Rb atoms which is prepared and trapped using an atom chip~\cite{Reichel11}. 
We initialize the non-equilibrium dynamics by transversally splitting this single 1D gas coherently into two nominally identical 1D systems, each containing half of the atoms, on average. Information about the total system is extracted using matter-wave interferometry between the two halves~\cite{Schumm05,Gring12,Langen13b,Kuhnert13}. This enables the time-resolved measurement of individual two-point and higher-order $N$-point phase correlation functions 
\begin{align}
C(&z_1,z_2,\ldots, z_N)\nonumber\\
&\sim\langle\Psi_1(z_1)\Psi_2^\dagger(z_1)\Psi_1^\dagger(z_2)\Psi_2(z_2)\cdots \Psi_1^\dagger(z_N)\Psi_2(z_N)\rangle\nonumber\\
&\sim\langle \exp[i\varphi(z_1)-i\varphi(z_2)+\cdots-i\varphi(z_N)]\rangle,
\end{align}
where $z_1, z_2, \ldots, z_N$ are $N$ coordinates along the length of the system, and $\varphi(z)$ the relative phase between the two halves (see Supplementary Materials). As we show in the following, these correlation functions reveal detailed information about the dynamics and the steady states of the system.

We start with the two-point correlation function $C(z_1,z_2)\sim\langle \exp[i\varphi(z_1)-i\varphi(z_2)]\rangle$. Previously, this correlation function was studied in regions where the system is approximately translationally invariant~\cite{Langen13b,Betz11}, i.e. $C(z_1,z_2)\equiv C(z_1-z_2)$. Here, more comprehensive information about generic many-body states is obtained by mapping the full correlation function $C(z_1,z_2)$ for any combination of the coordinates $z_1$ and $z_2$ (see Fig.~\ref{fig:Fig1}).  

%===========================================================================
\begin{figure}[tbp]
	\centering
		\includegraphics[width=0.45\textwidth]{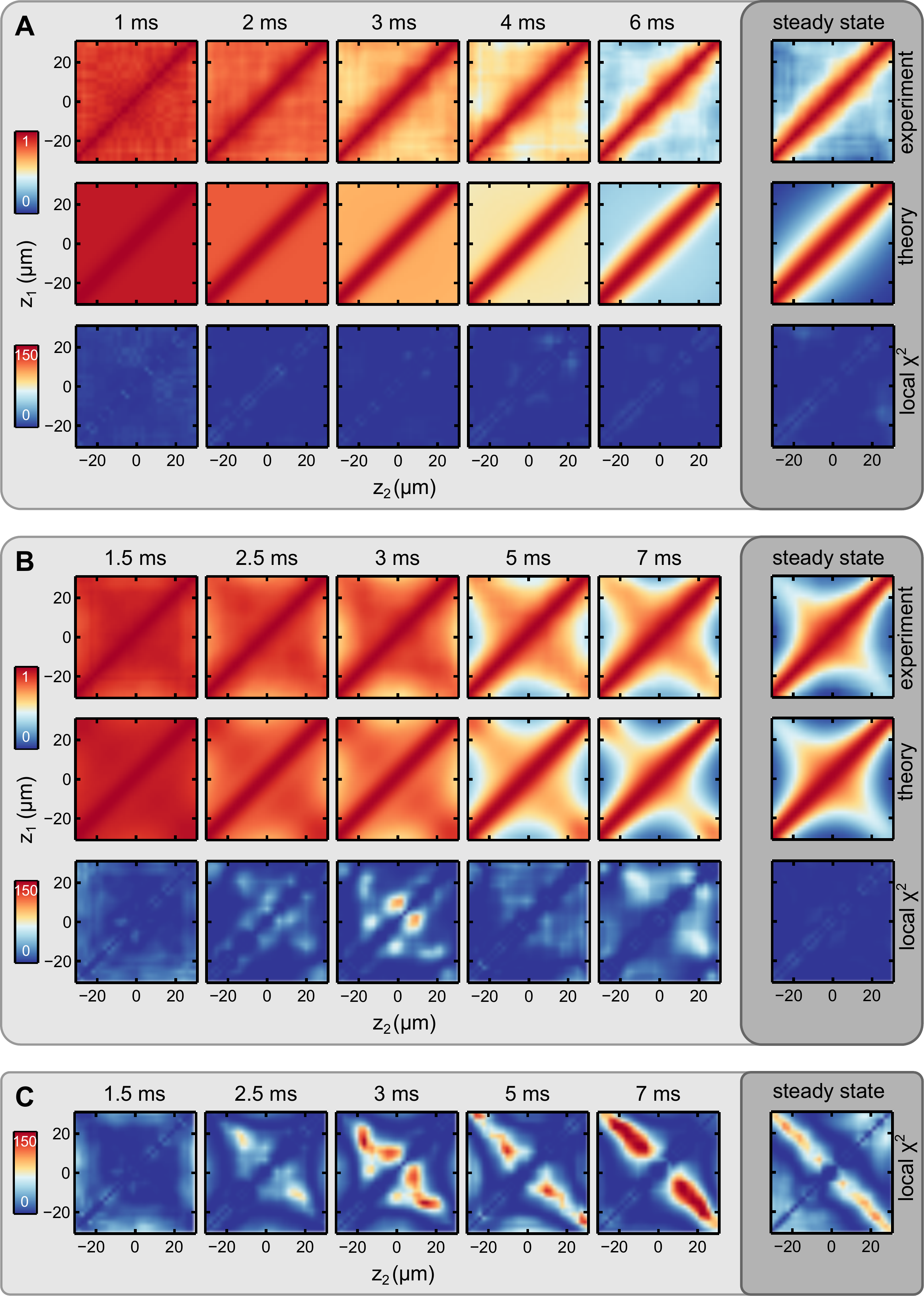}
	\caption{Two-point phase correlation functions $C(z_1,z_2)$ for increasing evolution time. Using two different splitting protocols, we prepare different initial states (\textbf{A},\textbf{B}). Both states show a characteristic maximum on the diagonal and a decay of correlations away from the diagonal.
	We quantify the agreement of our theoretical model and of the experiments using a $\chi^2$ analysis. The steady state and the dynamics in (\textbf{A}) can be well described by a single temperature $T_\mathrm{eff}$. As shown in (\textbf{C}), this single-temperature model fails for the steady state and the dynamics in (\textbf{B}), which require more temperatures to explain additional correlations on the anti-diagonal (see main text). The observation of different temperatures in the same steady state constitutes our observation of a GGE. The center of the system is located at $z_1=z_2=0$, color marks the amount of correlations between $0$ and $1$, and the local $\chi^2$ contribution between $0$ and $150$. The uncertainty of the correlation functions is estimated via bootstrapping over approximately $150$ experimental realizations (see Supplementary Information).}
	\label{fig:Fig2}
\end{figure}
%===========================================================================

Our observations following a typical splitting, which is fast compared to the dynamics of the system and therefore realizes a quench (see Supplementary Materials), are summarized in Fig.~\ref{fig:Fig2}A. As every point in the system is perfectly correlated with itself, the correlation functions exhibit a maximum on the diagonal $z_1=z_2$ for all times. Away from the diagonal, the system shows a light-cone-like decay of correlations~\cite{Langen13b} leading to a steady state. From a theoretical point of view, the emergence of this steady state is due to prethermalization~\cite{Berges:2004ce,Gasenzer:2005ze,Berges:2007ym,Eckstein09,Kitagawa11,Gring12}, which in the present case can be described as the dephasing of phononic excitations~\cite{Berges:2007ym,Bistritzer07,Kitagawa10,Kitagawa11}. The occupation numbers $n_{m}$ of these excitations are the conserved quantities of the corresponding integrable model (see Supplementary Materials). 

With a knowledge of the conserved quantities, we can directly calculate the Lagrange multipliers $\lambda_{m}$ for the GGE. In terms of the excitation energies $\epsilon_m$ they can be written as $\lambda_m =\beta_m\epsilon_m$ which defines an effective temperature $1/\beta_m$ for every excitation mode. 

For the steady state illustrated in Fig.~\ref{fig:Fig2}A the proportionality factor $\beta_m$ can be well described by $\beta_m\approx\beta_\mathrm{eff}=1/k_B T_\mathrm{eff}$. A fit yields $k_B T_\mathrm{eff}=(0.50\pm0.01)\times\mu$ which is independent of $m$ and in very good agreement with theory~\cite{Gring12,Kitagawa10}. Here, $\mu$ denotes the chemical potential in each half of the system. While being a GGE in principle, for our experiment which observes the relative phase between the two halves of the system, it becomes formally equivalent to the usual Gibbs ensemble with a single temperature $T_\mathrm{eff}$ (see Supplementary Materials).

To obtain direct experimental signatures of a genuine GGE, we modify the initial state so that it exhibits different temperatures for different excitation modes. This is accomplished by changing the ramp that splits the initial gas into two halves (see Supplementary Materials). The results are shown in Fig.~\ref{fig:Fig2}B. In addition to the maximum of correlations on the diagonal, we observe a pronounced second maximum on the anti-diagonal.  This corresponds to enhanced correlations of the points $z_1=-z_2$, which are located symmetrically around the center of the system. These correlations are a direct consequence of an increased (decreased) population of quasi-particle modes that are even (odd) under a mirror-reflection with respect to the longitudinal trap center. Consequently, the observations can be described, to a first approximation,  by the above theoretical model but with different temperatures, i.e. with $\beta_{2m}=1/[k_B(T_\mathrm{eff}+\Delta T)]$ for the even and $\beta_{2m-1}=1/[k_B(T_\mathrm{eff}-\Delta T)]$ for the odd modes, respectively. Fitting the experimental data of the steady state with this model we find $k_B T_\mathrm{eff}= (0.64\pm 0.01)\times\mu$, $\Delta T = (0.48\pm 0.01) \times T_\mathrm{eff}$ and a reduced $\chi^2\approx 6$. 

%-----------------------------------------------------------------------------------------
\begin{figure}[tb]
	\centering
		\includegraphics[width=0.45\textwidth]{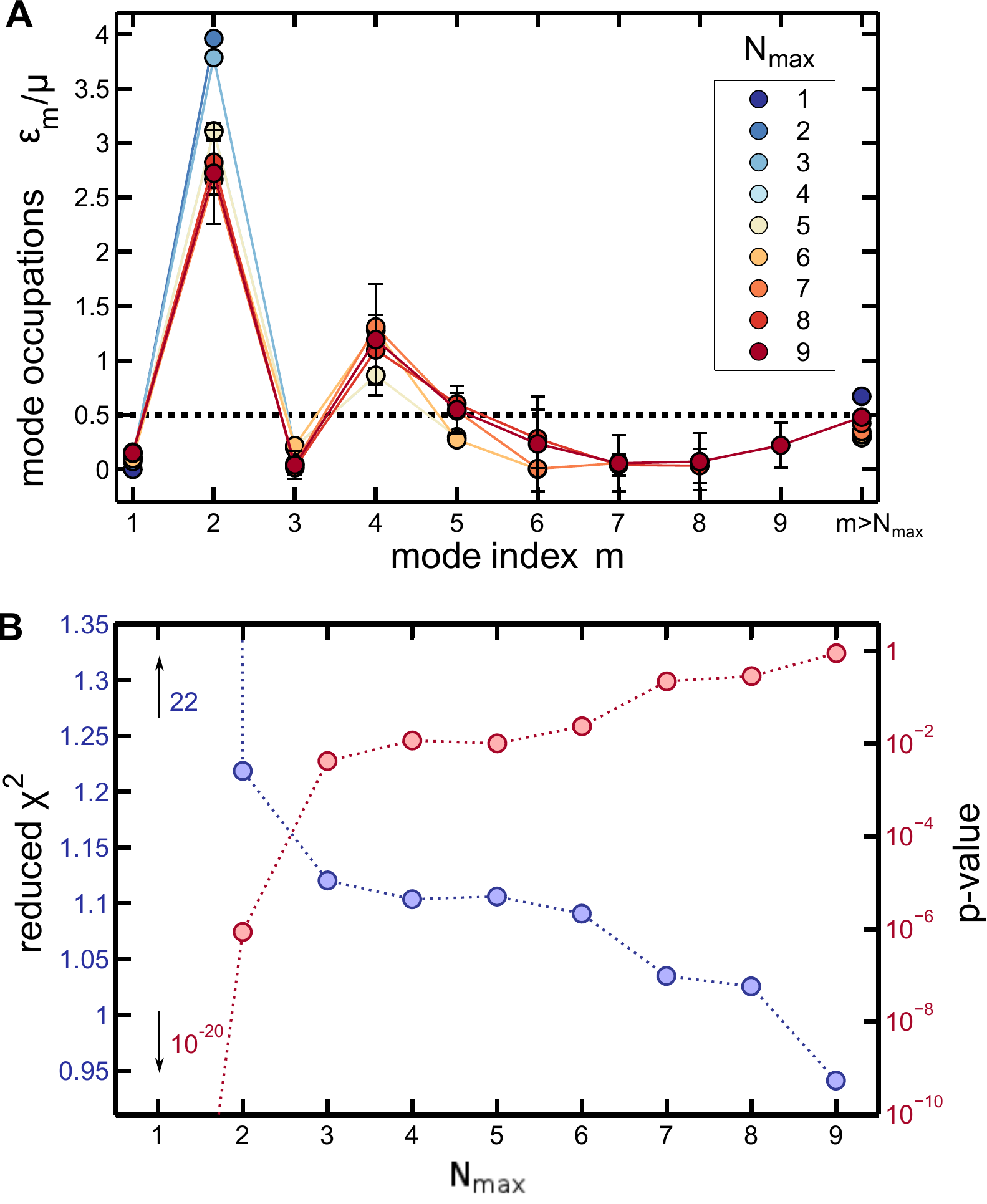}
	\caption{(\textbf{A}) Occupation numbers $n_m$ (in units of $\epsilon_m/\mu$) of the quasi-particle modes with index $m$, for different fitting procedures applied to the data from Fig.~\ref{fig:Fig2}B. The color of the points encodes results where modes up to $m=N_\mathrm{max}$ are fitted freely, while all higher modes are fit with the same occupation number. The plot clearly reveals how the occupation of the lowest even (odd) modes are increased (decreased) as compared to the single-temperature state from Fig.~\ref{fig:Fig2}A (dashed line).
	(\textbf{B}) Scaling of the reduced $\chi^2$ value, and of the corresponding p-value with $N_\mathrm{max}$. We observe that already the occupation numbers of the lowest $9$ modes and a single occupation number for all higher modes are sufficient to describe the experimental data to very good accuracy. This suggests that the observed GGE can be defined by only $10$ Lagrange multipliers, although there exists a much larger number of conserved quantities in the system.}
	\label{fig:Fig3}
\end{figure}
%-----------------------------------------------------------------------------------------

More detailed insights and a more accurate description of the experimental data can be gained by fitting the steady state with the individual mode occupations as free parameters. The results yield a reduced $\chi^2$ close to $1$ and thus a very good description of the experimental data  (see Fig.~\ref{fig:Fig2}B). As expected from our intuitive two-temperature model, the fitting confirms that the occupation of even modes is strongly enhanced, whereas the occupation of odd modes is reduced (see Fig.~\ref{fig:Fig3}A). Given these occupation numbers extracted from the steady state, our theoretical model also describes the complete dynamics very well. This clearly demonstrates that the different populations of the modes were imprinted onto the system by the splitting quench and are conserved during the dynamics. In contrast to that, a simple model based on a usual Gibbs ensemble with just one temperature for all modes clearly fails to describe the data (best fit: $T_\mathrm{eff}= (0.38\pm 0.01)\times\mu$, reduced $\chi^2\approx 25$), as visualized in Fig.~\ref{fig:Fig2}C. 

Notably, our fitting results for the GGE exhibit strong correlations between the different even modes and the different odd modes, respectively. This demonstrates the difficulty in fully and independently determining the parameters of such complex many-body states. In fact, any full tomography of all parameters would require exponentially many measurements. The results thus clearly show the presence of a GGE with at least two, but most likely many more temperatures.

Our work raises the interesting question how many Lagrange multipliers are needed in general to describe the steady state of a realistic integrable quantum system. Similar as in classical mechanics, where $N$ conserved quantities exist for a generic integrable system with $N$ degrees of freedom, integrability in quantum many-body systems has been proposed to be characterized by the fact that the number of independent local conserved quantities scales with the number of particles~\cite{Polkovnikov11}. Here we conjecture that most of the experimentally obtainable initial states evolve in time into steady states, which can be described to a reasonable precision by far less than $N$ Lagrange multipliers~\cite{Caux13,Barmettler2013}. This would have the appeal of a strong similarity to thermodynamics, where also only few parameters are needed to describe the properties of a system on macroscopic scales. 

To illustrate this in our specific case, we investigate in Fig.~\ref{fig:Fig3} how many distinct Lagrange multipliers need to be considered in the GGE to describe our data with multiple temperatures. Including more and more modes in the fitting of the experimental data, we find that the reduced $\chi^2$ values decrease and settle close to unity for $9$ included modes, with all higher modes being fitted by one additional Lagrange multiplier. 
Looking at the p-value for the measured $\chi^2$~\cite{Hughes} shows that only a limited number of Lagrange multipliers needs to be specified to describe the observables under study to the precision of the measurement. A simple numerical estimate based on the decreasing contribution of higher modes to the measured correlation functions and the limited imaging resolution leads to approximately $10$ Lagrange multipliers, which is in good agreement with our observations (see Supplementary Materials). Moreover, comparing this result with the single-temperature steady state that was discussed in the beginning illustrates how the complexity of the initial state plays an important role for the number of Lagrange multipliers that need to be included in a GGE.

%===========================================================================
\begin{figure}[ptb]
	\centering
		\includegraphics[width=0.45\textwidth]{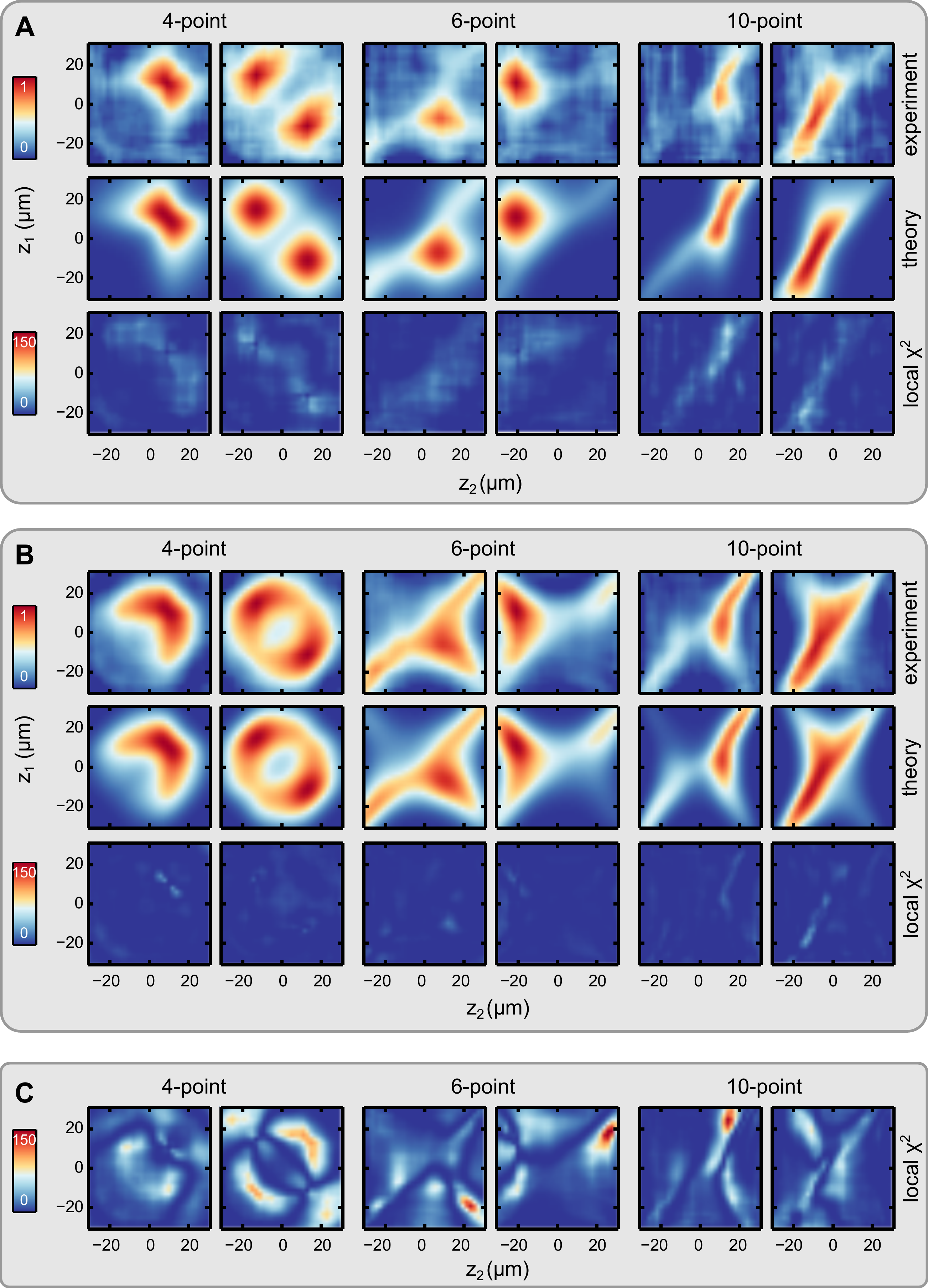}
	\caption{Examples of experimental 4-, 6-, and 10-point correlation functions reveal that the observed steady states also agree very well with the respective GGE predictions for more complex observables. In particular, differences between the steady state described by a single temperature (\textbf{A}) and the steady state described by multiple temperatures (\textbf{B}) are significant and can be very well captured by the theoretical model. (\textbf{C}) As for the two-point correlation functions, the single-temperature model can not describe the state with multiple temperatures. From left to right, coordinates are $C(z_1,10,z_2,10)$, $C(z_1,-12,z_2,14)$, $C(z_1,10,10,z_2,-20,10)$, $C(z_1,-8,8,z_2,-24,-20)$, $C(z_1,4,10,z_2,-8,z_2,-22,-18,10,-4)$ and $C(z_1,-22,-8,z_2,-22,-26,-22,z_2,-26,-24)$. All coordinates are given in $\mu$m and were randomly chosen to illustrate the high-dimensional data.}
	\label{fig:Fig4}
\end{figure}
%===========================================================================

In general, deviations of steady states from the GGE description are expected to manifest first in higher-order correlation functions. To provide further evidence for our theoretical description and the presence of a GGE, we show in Fig.~\ref{fig:Fig4} examples of measured four-point, six-point and ten-point correlation functions of the steady state. As the two-point correlation functions, they are in very good agreement with the theoretical model and clearly reveal the difference between the GGE and the usual Gibbs ensemble. This confirms that the description based on a GGE with the parameters extracted from the two-point correlation functions also describes many-body observables at least up to the $10^{\mathrm{th}}$ order.

In conclusion, we have observed direct experimental signatures for the emergence of a generalized Gibbs ensemble in the non-equilibrium evolution of an isolated quantum many-body system. This substantiates the importance of the maximum entropy principle and the generalized Gibbs ensemble as key aspects of the emergence of statistical mechanics from a microscopic unitary quantum evolution. We expect our measurements of correlation functions to high order to play an important role in new tomography techniques for complex quantum many-body states~\cite{Hubener13}. Moreover, the observed tuneability of the non-equilibrium states suggests that our splitting process could in the future be used to prepare states tailored for precision metrology~\cite{Berrada13}.

We acknowledge discussions with Eugene Demler, Emanuele Dalla Torre, Kartiek Agarwal, J\"urgen Berges, Markus Karl, Valentin Kasper, Isabelle Bouchoule, Marc Cheneau and Pjotrs Grisins. This work was supported by the EU (SIQS and ERC advanced grant QuantumRelax). We acknowledge support by the Austrian Science Fund (FWF) through the Doctoral Programme CoQuS (\textit{W1210}) (B.R. and T.S.), the Lise Meitner Programme M1423 (R.G.) and project P22590-N16 (I.E.M.). T.G. acknowledges support by the Deutsche Forschungsgemeinschaft (GA677/7,8), the University of Heidelberg (Center for Quantum Dynamics), and the Helmholtz Association (HA216/EMMI). This research was supported in part by the National Science Foundation under Grant No. NSF PHY11-25915. T.L., T.G., and J.S. thank the Kavli Institute for Theoretical Physics, Santa Barbara, for its hospitality.
%------------------------------------

\bibliography{biblio_new}

\begin{thebibliography}{10}

\bibitem{Shannon:1949}
C.~E. Shannon, W.~Weaver, {\it The mathematical theory of communication\/} (The
  University of Illinois Press, Urbana, IL, 1949).

\bibitem{Jaynes57}
E.~T. Jaynes, {\it Phys. Rev.\/} {\bf 106}, 620 (1957).

\bibitem{Jaynes57b}
E.~T. Jaynes, {\it Phys. Rev.\/} {\bf 108}, 171 (1957).

\bibitem{HuangBook}
K.~Huang, {\it Statistical mechanics\/} (Wiley, 1987).

\bibitem{Polkovnikov11}
A.~Polkovnikov, K.~Sengupta, A.~Silva, M.~Vengalattore, {\it Rev. Mod. Phys.\/}
  {\bf 83}, 863 (2011).

\bibitem{Gring12}
M.~Gring, {\it et~al.\/}, {\it Science\/} {\bf 337}, 1318 (2012).

\bibitem{Ronzheimer13}
J.~P. Ronzheimer, {\it et~al.\/}, {\it Phys. Rev. Lett.\/} {\bf 110}, 205301
  (2013).

\bibitem{Caux13}
J.-S. Caux, F.~H. Essler, {\it Phys. Rev. Lett.\/} {\bf 110}, 257203 (2013).

\bibitem{Calabrese2011}
P.~Calabrese, F.~H.~L. Essler, M.~Fagotti, {\it Phys. Rev. Lett.\/} {\bf 106},
  227203 (2011).

\bibitem{Kinoshita06}
T.~Kinoshita, T.~Wenger, D.~Weiss, {\it Nature\/} {\bf 440}, 900 (2006).

\bibitem{Rigol07}
M.~Rigol, V.~Dunjko, V.~Yurovsky, M.~Olshanii, {\it Phys. Rev. Lett.\/} {\bf
  98}, 050405 (2007).

\bibitem{Rigol2009a}
M.~Rigol, {\it Phys. Rev. Lett.\/} {\bf 103}, 100403 (2009).

\bibitem{Cazalilla06}
M.~A. Cazalilla, {\it Phys. Rev. Lett.\/} {\bf 97}, 156403 (2006).

\bibitem{Caux12}
J.-S. Caux, R.~M. Konik, {\it Phys. Rev. Lett.\/} {\bf 109}, 175301 (2012).

\bibitem{Rigol08}
M.~Rigol, V.~Dunjko, M.~Olshanii, {\it Nature\/} {\bf 452}, 854 (2008).

\bibitem{Kollar11}
M.~Kollar, F.~A. Wolf, M.~Eckstein, {\it Phys. Rev. B\/} {\bf 84}, 054304
  (2011).

\bibitem{Vosk2013}
R.~Vosk, E.~Altman, {\it Phys. Rev. Lett.\/} {\bf 110}, 067204 (2013).

\bibitem{Lieb63}
E.~H. Lieb, W.~Liniger, {\it Phys. Rev.\/} {\bf 130}, 1605 (1963).

\bibitem{Korepin93}
V.~Korepin, N.~Bogoliubov, A.~Izergin, {\it Quantum inverse scattering method
  and correlation functions\/} (Cambridge University Press, Cambridge, 1993).

\bibitem{Kuhnert13}
M.~Kuhnert, {\it et~al.\/}, {\it Phys. Rev. Lett.\/} {\bf 110}, 090405 (2013).

\bibitem{Langen13b}
T.~{Langen}, R.~Geiger, M.~{Kuhnert}, B.~{Rauer}, J.~Schmiedmayer, {\it Nature
  Physics\/} {\bf 9}, 640 (2013).

\bibitem{Petrov00}
D.~S. Petrov, G.~V. Shlyapnikov, J.~T.~M. Walraven, {\it Phys. Rev. Lett.\/}
  {\bf 85}, 3745 (2000).

\bibitem{Reichel11}
J.~Reichel, V.~Vuletic, eds., {\it Atom Chips\/} (Wiley, VCH, 2011).

\bibitem{Schumm05}
T.~Schumm, {\it et~al.\/}, {\it Nature Physics\/} {\bf 1}, 57 (2005).

\bibitem{Betz11}
T.~Betz, {\it et~al.\/}, {\it Phys. Rev. Lett.\/} {\bf 106}, 020407 (2011).

\bibitem{Berges:2004ce}
J.~Berges, S.~Borsanyi, C.~Wetterich, {\it Phys. Rev. Lett.\/} {\bf 93}, 142002
  (2004).

\bibitem{Gasenzer:2005ze}
T.~Gasenzer, J.~Berges, M.~G. Schmidt, M.~Seco, {\it Phys. Rev. A\/} {\bf 72},
  063604 (2005).

\bibitem{Berges:2007ym}
J.~Berges, T.~Gasenzer, {\it Phys. Rev. A\/} {\bf 76}, 033604 (2007).

\bibitem{Eckstein09}
M.~Eckstein, M.~Kollar, P.~Werner, {\it Phys. Rev. Lett.\/} {\bf 103}, 056403
  (2009).

\bibitem{Kitagawa11}
T.~Kitagawa, A.~Imambekov, J.~Schmiedmayer, E.~Demler, {\it New J. Phys.\/}
  {\bf 13}, 073018 (2011).

\bibitem{Bistritzer07}
R.~Bistritzer, E.~Altman, {\it PNAS\/} {\bf 104}, 9955 (2007).

\bibitem{Kitagawa10}
T.~Kitagawa, {\it et~al.\/}, {\it Phys. Rev. Lett.\/} {\bf 104}, 255302 (2010).

\bibitem{Barmettler2013}
P.~{Barmettler}, C.~{Kollath}, {\it arxiv:1312.5757\/}  (2013).

\bibitem{Hughes}
I.~G. Hughes, T.~Hase, {\it Measurements and their Uncertainties\/} (Oxford
  Univ. Press, 2010).

\bibitem{Hubener13}
R.~H\"ubener, A.~Mari, J.~Eisert, {\it Phys. Rev. Lett.\/} {\bf 110}, 040401
  (2013).

\bibitem{Berrada13}
T.~Berrada, {\it et~al.\/}, {\it Nature Comm.\/} {\bf 4}, 2077 (2013).

\bibitem{Lesanovsky06}
I.~Lesanovsky, {\it et~al.\/}, {\it Phys. Rev. A\/} {\bf 73}, 033619 (2006).

\bibitem{Smith13}
D.~{Adu Smith}, {\it et~al.\/}, {\it New J. Phys.\/} {\bf 15}, 075011 (2013).

\bibitem{Whitlock03}
N.~K. Whitlock, I.~Bouchoule, {\it Phys. Rev. A\/} {\bf 68}, 053609 (2003).

\bibitem{Geiger2014}
R.~Geiger, T.~Langen, I.~Mazets, J.~Schmiedmayer, {\it New J. Phys.\/} {\bf
  16}, 053034 (2014).

\bibitem{Meyer1990}
H.-D. Meyer, U.~Manthe, L.~S. Cederbaum, {\it Chem. Phys. Lett.\/} {\bf 165},
  73 (1990).

\bibitem{Alon08}
O.~E. Alon, A.~I. Streltsov, L.~S. Cederbaum, {\it Phys. Rev. A\/} {\bf 77},
  033613 (2008).

\bibitem{Grond09}
J.~Grond, J.~Schmiedmayer, U.~Hohenester, {\it Phys. Rev. A\/} {\bf 79}, 021603
  (2009).

\end{thebibliography}

\bibliographystyle{Science}

\cleardoublepage

\section*{Supplementary Materials}

\section{Preparation of the 1D gas}
The initial one-dimensional (1D) Bose gas is prepared using our standard procedure to produce ultracold gases of ${}^{87}$Rb on an atom chip~\cite{Reichel11}. The initial trap has measured harmonic frequencies of $\omega_\perp=2\pi\times(2.1\pm 0.1)\,$kHz in the radial direction and $\omega_z=2\pi\times(10\pm 0.5)\,$Hz in the longitudinal direction. The temperature, atom number and chemical potential are $T=30-110\,$nK, $N=(5000\pm 500)$ and $\mu=\hbar\times (1.3\pm 0.1)\,$kHz, respectively, such that the 1D condition $\mu, k_B T\leq \hbar \omega_\perp$ is well fulfilled. The gas typically has a length of $100\,\mu$m, from which we use the central $60\,\mu$m for our analysis.

\section{Splitting}
To split the initial gas, the static trapping potential is superimposed with linearly polarized radio-frequency (RF) radiation which is applied by means of wires on the atom chip~\cite{Schumm05}. The RF frequency of $365\,$kHz corresponds to a detuning of $30\,$kHz to the red of the $m_F=2\rightarrow m_F=1$ transition. Applying the RF adiabatically transfers the atoms into dressed states. This leads to a smooth transformation of the static harmonic trap into a dressed-state double-well potential~\cite{Lesanovsky06} which is aligned perpendicular to gravity. The control parameter for this deformation and the properties of the final double-well potential is given by the amplitude of the RF radiation. Controlling this amplitude thus enables the realization of different splitting protocols. For the data presented in Fig.~2A of the main text, we linearly increase the RF amplitude to $8\,$mA over a time of $30\,$ms. This is followed by a faster increase to $25\,$mA in $12\,$ms. The excitation of a breathing mode is intrinsic to this splitting procedure, due to the halving of the atom number. It can be neglected on the timescale of the experiments presented in this work. For the data presented in Fig.~2B of the main text, the RF amplitude is increased linearly to the final $25\,$mA within a single $17\,$ms long ramp. For both RF ramp protocols the rapid decoupling of the two gases happens approximately $3\,$ms before the end of the RF amplitude ramp and within a period of less than $500\,\mu$s. After the splitting, the tunnel coupling between the two gases is negligible. The final double well has measured trap frequencies of $\omega_\perp=2\pi\times (1.4\pm 0.1)\,$kHz in the radial direction and $\omega_z=2\pi\times (7.5\pm 0.5)\,$Hz in the longitudinal direction. A barrier height of $6\,$kHz is estimated from simulations of the double-well potential. 
\\

\section{Measurement of the phase correlation functions}
To extract the phase correlation functions we record the resulting matter-wave interference pattern of the two 1D Bose gases after $15.9\,$ms time-of-flight expansion using standard absorption imaging~\cite{Gring12,Smith13}. The Gaussian point-spread function (PSF) of the imaging system has a measured $1/\sqrt{e}$ radius of $3.8\,\mu$m, which can be taken into account in the theoretical description by applying a convolution to the phase variances $\langle \varphi(z) \varphi(z)' \rangle$. 

The local position of the fringes in the fluctuating interference pattern directly corresponds to the relative phase $\varphi(z)$ between the two gases. We extract this relative phase by fitting a sinusoidal function to each pixel line in the interference pattern~\cite{Langen13b}. 

On the theoretical side, we define the even-order many-body correlation functions as
\begin{widetext}
\begin{align}
	C(z_1,z_2) &=\frac{\langle \Psi_1(z_1)\Psi_2^\dagger(z_1)\Psi_1^\dagger(z_2)\Psi_2(z_2)\rangle}{\langle |\Psi_1(z_1)|^2\rangle \langle |\Psi_2(z_2)|^2\rangle},\nonumber\\
	C(z_1,z_2,z_3,z_4) &= \frac{\langle \Psi_1(z_1)\Psi_2^\dagger(z_1)\Psi_1^\dagger(z_2)\Psi_2(z_2) \Psi_1(z_3)\Psi_2^\dagger(z_3)\Psi_1^\dagger(z_4)\Psi_2(z_4)\rangle}{\langle |\Psi_1(z_1)|^2\rangle \langle |\Psi_1(z_2)|^2\rangle \langle |\Psi_2(z_3)|^2\rangle\langle |\Psi_2(z_4)|^2\rangle},\\
	&\,\,\,\vdots\nonumber
\end{align}
\end{widetext}

Note that odd-order correlation functions for our system decay exponentially with time and thus do not contain further information about the steady states. Following the standard procedure we expand the bosonic fields  in terms of density and phase, $\Psi_{j}(z) = \mathrm{exp}\{i\,\theta_{j}(z)\} \sqrt{\rho_{j}(z)}$, and consider small fluctuations around equal equilibrium densities of the two clouds, ${\rho}_{j}(z) = \rho_{0}(z) + \delta{\rho}_{j}(z)$. Here, the index $j=1,2$ labels the two individual gases. Writing the local relative phase as $\varphi(z)=\theta_1(z)-\theta_2(z)$ and neglecting the fluctuations $\delta{\rho}_{j}(z)$, we find the expressions
\begin{widetext}
\begin{align}
	C(z_1,z_2) &\approx \langle \exp[i\varphi(z_1)-i\varphi(z_2)]\rangle,\nonumber\\
	C(z_1,z_2,z_3,z_4) &\approx \langle \exp[i\varphi(z_1)-i\varphi(z_2)+i\varphi(z_3)-i\varphi(z_4)]\rangle,\\
	&\,\,\,\vdots\nonumber
\end{align}
\end{widetext}
for the phase correlation functions. They only contain the relative phase $\varphi(z)$ and can thus be directly calculated from the experimental data. In this procedure, the expectation value is evaluated by averaging over many experimental realizations. The corresponding uncertainty is estimated using a bootstrapping method~\cite{Hughes}. To check our theoretical description (see below), we calculate the reduced $\chi^2$ values. Moreover, comparing the obtained results to a $\chi^2$ distribution we deduce a p-value or statistical significance~\cite{Hughes}, i.e. a probability that our model produces a $\chi^2$ at least as big as the one calculated from the experimental data (shown in Fig.~\ref{fig:Fig3}B).

Note that only the phase correlation function, but not the integrated interference contrast as studied in our previous work~\cite{Gring12,Kuhnert13}, is sensitive enough to reveal the subtle differences between the two splitting protocols. The reason for this is that the effect of the two different temperatures on the contrast is only a small offset, which is negligible for the given experimental precision. Studying the contrast one is thus not able to observe the differences in the splitting process. 

%==========================================================================
\section{Prethermalization to a generalized Gibbs ensemble}
\label{sec:GGEcalculations}
In this section, we provide more details of the theoretical model and the procedure for extracting the parameters of the GGE from our data. Further details will be given in a future work.

%-----------------------------------------------------------------------------------------
\subsection{Split one-dimensional condensate}
Two coupled 1D degenerate Bose gases can be described by the Hamiltonian
\begin{align}
 H &= \int \mathrm{d} z~\Big[\Psi_{1}^{\dagger}\Big( -\frac{\hbar^2\partial_{z}^{2}}{2m}+  U(z) - \mu  + \frac{g}{2} \Psi_{1}^{\dagger}  \Psi_{1}^{}\Big)\Psi_{1}^{} 
 \nonumber\\
& \qquad- J \Psi_{1}^{\dagger} \Psi_{2}^{}  + (1 \leftrightarrow 2 )\Big] ~ \mathrm{,}
\end{align}
with longitudinal trapping potential $U$, chemical potential $\mu$, interaction constant $g$, and tunneling coupling $J$.
Transverse excitations are frozen out, i.e., $\Psi_{j}\equiv\Psi_{j}(z)$. 

This system can be written in terms of symmetric and anti-symmetric degrees of freedom, corresponding to the combinations  $\delta\rho_\mathrm{s} = \delta\rho_1 + \delta\rho_2$,  $\varphi_\mathrm{s} = (\theta_1 + \theta_2)/{2}$, and $\delta\rho_\mathrm{a} = ({\delta\rho_1 - \delta\rho_2})/{2}$, $\varphi_\mathrm{a} = \theta_1 - \theta_2$. 
%Here, we have replaced the label for the relative phase $\varphi\equiv\theta_\mathrm{a}$ to simplify our notation in the following. 
Within the range of energy scales realized in the experiment, the system is described by two sets of Bogoliubov-de Gennes equations \cite{Whitlock03}. To second order in the small fields $\delta\rho_{\mathrm{s,a}}$, $\varphi_{\mathrm{s,a}}$, the Hamiltonian separates into symmetric and anti-symmetric parts and can be diagonalized within a quasi-particle basis,
\begin{equation}
\label{eq:effective free hamiltonian}
H^{(2)} = \sum_{{\alpha=\mathrm{a,s}}\atop m} \epsilon_{\alpha m}\, \beta_{\alpha m}^{\dagger} \beta_{\alpha m} ~\text{.}
\end{equation}
The Hamiltonian $H^{(2)}$ is integrable and describes a set of uncoupled harmonic oscillators with energies $\epsilon_{\alpha m}$. The conserved quantities are given by the occupation numbers $n_m=\langle\beta_{\alpha m}^{\dagger} \beta_{\alpha m}\rangle$ of the individual oscillators, where $\beta_{\alpha m}^{\dagger}$ and $\beta_{\alpha m}$ are bosonic creation and annihilation operators. 

In the splitting process, the symmetric degrees of freedom inherit all the thermal energy of the initial state and are therefore trivially described by an additional temperature~\cite{Kitagawa11,Gring12}. 

In the experiment, only the anti-symmetric degrees of freedom, $\delta\rho_\mathrm{a}$, $\varphi_\mathrm{a}$, are probed via the matter-wave interference patterns. 
We therefore omit, in the following and in the main text, the index ``a" to the respective operators and frequencies to simplify the notation. 

When describing the fluctuations properties of the anti-symmetric degrees of freedom, we take into account that the splitting process is performed on a time scale which is small compared to the speed of sound in the system. Thus, we can neglect non-local correlations. For every point along $z$, the probability of an atom to go to the right or the left gas is essentially the same, such that, immediately after the division of the cloud into two halves, one has \cite{Kuhnert13,Kitagawa11}
\begin{align}
\label{eq:fluctuations}
\langle\delta\rho(z_1)\delta\rho(z_2)\rangle 
&= \frac{\rho_{0}(z_1)}{2} \delta(z_1-z_2) \,,
\nonumber\\
\langle\varphi(z_1)\varphi(z_2)\rangle
&= \frac{1}{2 \rho_{0}(z_1)} \delta(z_1-z_2) ~\text{.}
\end{align}
At the edges of the cloud where the density $\rho_{0}$ vanishes, the local fluctuations of density and phase saturate to the respective values for a vacuum state. 

%-----------------------------------------------------------------------------------------
\subsection{The generalized Gibbs ensemble} 
During the prethermalization period, the dynamics of the system can be well modeled as a dephasing process of the initially excited, nearly free quasi-particle modes \cite{Bistritzer07,Kitagawa10,Kitagawa11,Berges:2004ce}. 
These quasi-particle modes are defined by means of the Bogoliubov expansion 
\begin{align}
b(z)=\sum_{m}[u_{m}(z)\exp(-i\epsilon_{ m}t)\beta_{ m}+v_{m}^{*}(z)\exp(i\epsilon_{ m}t)\beta_{ m}^{\dagger}]
\end{align}
of the quadrature field $b(z)= \varphi(z)\sqrt{\rho_0/2} - i{\delta\rho(z)}/\sqrt{2\rho_0}$.
This expansion involves the mode functions $u_{m}$, $v_{m}$ and the Fock operators which obey $[\beta_{ n},\beta_{ m}^{\dagger}]=\delta_{nm}$, $[\beta_{ n},\beta_{ m}]=0$. 

In the prethermalized state, detailed information about the initial conditions such as the occupation numbers of the quasi-particle modes are not yet lost. 
In contrast, bulk observables like the equation of state already assume their final, thermalized values \cite{Berges:2004ce}.
For a Bose quasi-condensate, this implies that equipartition between kinetic and interaction energies is reached to a good approximation \cite{Gasenzer:2005ze}, from which a kinetic temperature can be defined.
%Hence, the conserved quantities, to a good approximation, are described by the quasi-particle number operators $\hat{\mathcal{I}}_{m}=\hat{\beta}^{\dagger}_{m}\hat{\beta}_{m}$. 
Given the set of conserved quasi-particle occupation numbers $\hat{\mathcal{I}}_{m}={\beta}^{\dagger}_{ m}{\beta}_{ m}$, a maximum-entropy state takes the form of the GGE in Eq.\eq{gge}, with the Lagrange multipliers $\lambda_{m}$ adjusted to yield the time-invariant expectation values $\mathcal{I}_{m}=\langle\hat{\mathcal{I}}_{m}\rangle\equiv n_m$ \cite{Rigol07,Jaynes57}.
These values are the result of the particular initial quench, i.e., depend on the details of the splitting process.
Due to the bosonic nature of the excitations, the partition function $Z$ of the GGE can be written, using the quasi-particle Fock basis, as
\begin{equation}
  Z = \prod_{m} \sum_{n_m\ge0} e^{- \lambda_m n_m} = \prod_{m} \frac{1}{1-e^{-\lambda_m}} ~\text{.}
\end{equation}
Inverting the expression for the average occupations $\mathcal{I}_{m}=-Z^{-1}\partial_{\lambda_{m}}Z$ gives the Lagrange multipliers
\begin{equation}
  \lambda_{m} = \ln\left(1+\mathcal{I}_{m}^{-1}\right) ~\text{.}
\end{equation}
Writing the multipliers in terms of the mode energies, $\lambda_m \equiv \beta_m \epsilon_{ m}$, the GGE defines different effective temperatures for every mode, $\beta_{m}=(k_\mathrm{B}T_{m})^{-1}$. 
The GGE reduces to a Gibbs ensemble if the $\lambda_{m}$ are proportional to the energies, $\lambda_m = \beta_\mathrm{eff} \epsilon_{ m}$, with an $m$-independent proportionality parameter $\beta_\mathrm{eff}$. 
Hence, if $\beta_m \approx \beta_{\text{eff}} \equiv   (k_{\text{B}} T_\mathrm{eff})^{-1}$, within some range of energies, the in any case prethermalized system is well described, within that range, by a Gibbs ensemble with a single effective temperature $T_{\text{eff}}$. The observation of such a system is demonstrated in Fig.~\ref{fig:Fig2}A.

Extending upon this, the data shown in Fig.~\ref{fig:Fig2}B reflects a more general prethermalized gas, described by a GGE with more temperatures. In the following, we summarize the procedure for obtaining the relevant effective mode temperatures for the GGE.

%-----------------------------------------------------------------------------------------
\subsection{Homogeneous case}
In absence of a trapping potential, $U(z) = 0$, the anti-symmetric quasi-particle excitations on the background of a uniform density $\rho_{0}(z) = n_0$ are characterized by the solutions
$f_{k}^{\pm}(z) = u_{k}\pm v_{k}={L}^{-1/2} ({\epsilon_k}/{E_k})^{\pm1/2}\, e^{i k z}$, with wave-number $k$ dependent mode frequencies $\epsilon_k = \sqrt{(E_k + 2J) [(E_k + 2J) + 2mc_{s}^{2}]}$, where $E_k = {\hbar^2 k^2}/({2m})$ and $c_{s} = \sqrt{g n_0/m}$.\\
Inserting these, for the case of perfect splitting ($J=0$) into the inverted Bogoliubov expansion, and neglecting the contribution of the highly suppressed phase fluctuations, gives mean quasi-particle occupations $n_k = \langle \beta_{k}^{\dagger} \beta_{k}\rangle = {\epsilon_k}/({4 E_k})$.
The corresponding Lagrange multipliers are $\lambda_{k}\simeq 2\epsilon_{k}/\mu$, $\mu=mc_{s}^{2}$, in the sound regime of wave numbers $\hbar k\lesssim mc_{s}$.
This shows that in the regime where the dispersion is linear in $k$, the GGE is equivalent to a Gibbs ensemble.
The system is prethermalized in the anti-symmetric degrees of freedom, with an effective temperature $\beta_\mathrm{eff}^{-1}=\mu/2$.
Note that, for wave numbers $\hbar k\gg mc_{s}$, the resulting Lagrange multipliers become $k$-independent and can be interpreted as an effective chemical potential for those modes.
Within the transition regime $\hbar k\simeq mc_{s}$, the non-trivially $k$-dependent $\lambda_{k}$ lead beyond a Gibbs description.
%This should, however, not lead to the false conclusion of a diverging energy as a real splitting process is bound to inject only a finite amount of energy.

%-----------------------------------------------------------------------------------------
\subsection{Harmonically trapped condensates}
To take into account the longitudinal harmonic confinement in the trap, $U(z)=  m \omega^2 z^2/2$,
we use the solutions of the Bogoliubov equations in the Thomas-Fermi approximation of the condensate \cite{Petrov00,Geiger2014},
\begin{equation}
f_{m}^{\pm}(x) = \sqrt{\frac{m+\frac{1}{2}}{R_{\text{TF}}}} \bigg[ \frac{2 \mu}{\epsilon_m} (1-x^2) \bigg]^{\pm {1}/{2}} P_m(x) ~\text{,}
\end{equation} 
with $n_{0}=\mu/g$, $R_\mathrm{TF}=(2\mu/m)^{1/2}\omega^{-1}$, and $x = z / R_{\text{TF}}$.
$P_m$ are the Legendre polynomials of order $m$, and the mode energies read $\epsilon_m = \hbar\omega_m=\hbar \omega \sqrt{m(m+1) / 2}$. Assuming homogeneous density fluctuations $\rho_{0}(z)=n_{0}$ in \Eq{fluctuations} and neglecting again the highly suppressed phase fluctuations, one finds the quasi-particle occupations $n_{m}= \langle \beta_{m}^{\dagger} \beta_{m}\rangle=\mu/ 2 \epsilon_{m}$. 
With this, the Lagrange multipliers, for $\epsilon_m \lesssim \mu/2$, result as $\lambda_{m} = {2}\epsilon_{m}/\mu$, giving again a single mode temperature $\beta_{\mathrm{eff}}^{-1}=\mu/{2}$. At higher energies the expansion of the logarithm is no longer valid, and the GGE deviates from the special Gibbs case. Nevertheless, due to the low occupation of the higher modes, the thermal approximation holds for a wide range of experimental parameters.\\
Using the above approximations the time-evolution of the correlations is obtained as
\begin{equation}
C(z_1,z_2,t) = \langle e^{i \varphi(z_1,t) - i \varphi(z_2,t)} \rangle = e^{ - \frac{1}{2} \langle \delta \varphi(t)^2 \rangle_{z_1z_2}}
\end{equation}
with the phase variance
\begin{align}
\langle \delta &\varphi(t)^2 \rangle_{z_1z_2}\equiv\langle [\varphi(z_1,t)-\varphi(z_2,t)]^2\rangle\nonumber\\
=  &\sum_{m>0} \frac{g (m+\frac{1}{2})}{R_{\text{TF}} \epsilon_m} 2 \sin^2(\omega_m t) \Big[ P_m(x_1) - P_m(x_2) \Big]^2 2 n_m  ~\text{.}
\end{align}
Similar results can be obtained for all higher even-order correlation functions. The contributions to this sum strongly decrease with increasing $m$. Taking into account the optical resolution leads to the estimate that modes with $m\gtrsim 10$ play a negligible role in the description of the observed correlation functions, which we find to be in good agreement with the experimental results.

\subsection{Imbalanced occupation numbers}
To obtain the occupation numbers $n_{m}$ of the excitation modes, we use two different models. The simplest choice is to assume different occupation numbers for the even and odd modes,
\begin{align}
\label{eq:imbalQPoccupnos}
n_{2m}&=\left[\beta_\mathrm{eff}^{-1}+\tilde\beta^{-1}\right]/\epsilon_{2m},
\nonumber\\
n_{2m-1}&=\left[\beta_\mathrm{eff}^{-1}-\tilde\beta^{-1}\right]/\epsilon_{2m-1}.
\end{align}
Hence, different effective temperatures apply for the two different sets of modes.

A more refined ansatz is to freely fit the occupations of the lowest $N_\mathrm{max}$ modes, $n_{m}$, with $m=1,\ldots,N_\mathrm{max}$. Additionally, all higher modes with $m>N_\mathrm{max}$ are all fitted with the same effective temperature. In Fig.~\ref{fig:Fig3} of the main text we show that only the first $9$ lowest lying modes have to be included to describe our experimental observations. Contributions of higher modes to the phase correlation function are highly suppressed. Moreover, as Fig.~\ref{fig:Fig3} readily shows, few-mode excitations alone are not able to capture the experimental observations.  

Note that we have experimentally and theoretically excluded mean-field effects such as mean 
particle number imbalance, different trap frequencies in the two wells, or long-lived coherences between 
Bogoliubov modes, arising for inhomogeneous initial density fluctuations proportional to $\rho_{0}(z)$, as 
possible explanations for the different occupation numbers. To exclude that experimental shot-to-shot fluctuations of the occupation numbers are the reason for the observed structure of the correlation functions, we have simulated such fluctuations numerically. We find that this cannot explain the observations.   

To understand the microscopic origin of the multiple observed temperatures, it is essential to understand the exact physics that occur during the splitting process. In the most simple picture an infinitely fast splitting process can be understood as a binomial distribution of the atoms into the two halves of the system, as described by Eq.~6. This simple picture leads to the emergence of a single temperature. However, it has previously been shown that complex non-linearities can appear in realistic splitting protocols. They result in the creation of squeezed states where the binomial atom number fluctuations predicted by the simple model are strongly reduced~\cite{Berrada13}. Our observation of multiple temperatures could be explained along these lines using locally reduced atom number fluctuations at the edges of the cloud. 

Due to the non-linearities in the splitting process, the resulting local atom number fluctuations have to be determined numerically. In this context, the splitting process has previously been simulated using classical field methods or (Multi-Layer) Multiconfigurational Time-Dependent Hartree for Bosons \mbox{(MCTDHB)}~\cite{Meyer1990,Alon08,Grond09}. However, a full theoretical model including the longitudinal degree of freedom has so far remained elusive. While classical fields can only account for thermal fluctuations using stochastic methods, MCTDHB has been used to study the creation of squeezing during the splitting of gases containing up to $\mathcal{O}(100)$ particles in zero dimensions. A modeling of the experimental splitting process involving the 1D direction and thousands of particles would require significantly more self-consistent orbitals and is thus far beyond reach of current computational resources. Consequently, we understand our measurements as an important benchmark for future simulations.

\end{document}